\documentclass[prl,showpacs,twocolumn]{revtex4}

\usepackage{epsfig}
\usepackage{color}

\usepackage{graphicx}

\usepackage{amssymb,amsfonts,amsmath}

\begin{document}

\title{Selective advantage of diffusing faster}

\author{Simone Pigolotti$^{1}$ and Roberto Benzi$^2$}
\affiliation{$^1$Dept. de Fisica i Eng. Nuclear, Universitat
  Politecnica de Catalunya Edif. GAIA, Rambla Sant Nebridi 22, 08222
  Terrassa, Barcelona, Spain. $^2$Dipartimento di Fisica, Universita'
  di Roma ``Tor Vergata'' and INFN, via della Ricerca Scientifica 1,
  00133 Roma, Italy. }

\date{today}

\begin{abstract} 
  We study a stochastic spatial model of biological competition in
  which two species have the same birth and death rates, but different
  diffusion constants. In the absence of this difference, the model
  can be considered as an off-lattice version of the Voter model and
  presents similar coarsening properties. We show that even a relative
  difference in diffusivity on the order of a few percent may lead to
  a strong bias in the coarsening process favoring the more agile
  species. We theoretically quantify this selective advantage and
  present analytical formulas for the average growth of the fastest
  species and its fixation probability.
\end{abstract}


\maketitle

Different physical, social and biological systems can be described by
models belonging to the Voter Model (VM) universality class \cite{dornic}. An
important example in biology is the
neutral Stepping Stone Model \cite{6,7} whose dynamics explains
qualitative and quantitatively spreading and fixation of competing
populations on a Petri dish \cite{hall_exp,korolev_sweeps}. In
statistical physics, the VM is characterized by the existence of two
symmetric absorbing states and a coarsening process without surface
tension. Its macroscopic dynamics corresponds to the Langevin equation
\begin{equation}
\label{VM}
\partial_t f(\mathbf{x},t) = D \Delta f + \sqrt{\Gamma f(1-f)} \eta(\mathbf{x},t)
\end{equation}
where $D$ is the diffusion constant, $\Gamma$ the noise amplitude and
$\eta(\mathbf{x},t)$ is a {\it $\delta$ }-correlated white noise. In
biological applications, the field $f$ usually represents the
frequency of an allele, i.e., the local fraction of individuals
carrying a given mutation. When mutants have a "selective advantage"
$s$ over the wild type (i.e. a difference in reproduction rate),
an additional term appears in Eq. (\ref{VM}), which becomes a
stochastic version of the celebrated
Fisher-Kolmogorov-Petrovskii-Piscounov (FKPP) equation
\begin{equation}
\label{FKPP}
\partial_t f (\mathbf{x},t)= D \Delta f + s f(1 -f)+ \sqrt{\Gamma f(1-f)}\eta(\mathbf{x},t)
\end{equation}
In the absence of noise, Eq. (\ref{FKPP}) predicts a
well-defined range expansion velocity of the mutants, $v=2\sqrt{Ds}$
\cite{fisher,kolmogorov}. The same result is valid in the presence of
weak multiplicative noise, up to logarithmic corrections
\cite{derrida,pechenik}, while in the strong noise regime one should
expect a difference expression for the velocity
\cite{doering,hallat_strongnoise}.

The stochastic FKPP equation is radically different from the VM. One
can think of $s$ in a similar way as the external field in the Ising
model, breaking the $f\leftrightarrow (1-f)$ symmetry and thus driving
the system away from the critical point, which is recovered for
$s=0$. For $s>0$, Eq.(\ref{FKPP}) predicts $f=1$ to be the
deterministic asymptotic stable equilibrium of the system, the state
$f=0$ being unstable.  Often, the critical behaviors of Langevin
equation such as variants of Eq. (\ref{VM}) can be understood by
analyzing their corresponding deterministic dynamics, either by means
of its associated mean field potential \cite{hammal} or by the
Hamiltonian dynamics obtained by the path integral formulations
\cite{elgart}.

Due to the relevance of the VM in non-equilibrium phenomena, it is
interesting to understand whether there exists more general, possibly
non-deterministic mechanisms to break the VM universality class. In
biological terms, this amounts to ask whether an effective
selective advantage can be achieved without any asymmetry in the birth
and death rates. For example, it has been recently shown
\cite{hall_noise} that an asymmetry in the carrying capacity (i.e. the
global biological mass) of the two alleles can induce an effective
selective advantage.

In this Letter, we show that an effective selective advantage emerges
in a competition model between two species diffusing at different
speeds, but otherwise neutral.  In biology, this setting is relevant
to assess the evolutionary importance of movement, for example in
species which exist in motile and non-motile variants, such as
bacteria with and without flagellum. We will show that, in this case,
competition is biased towards the fastest species. This is
equivalent to an effective selective advantage which depends both on
noise and spatial fluctuations, and is proportional to both the noise
amplitude and the difference in diffusivity.

\begin{figure}[htb]
\begin{center}
\includegraphics[width=9cm]{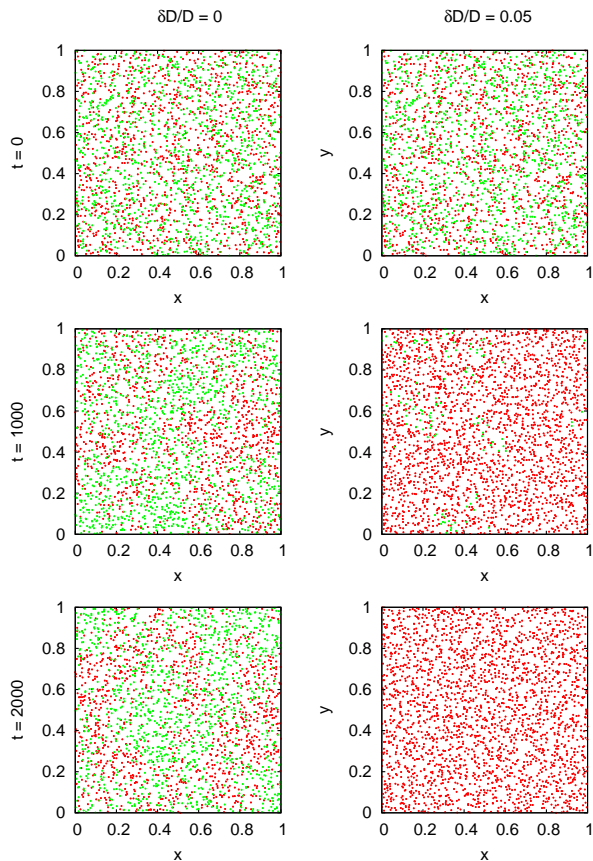}
\caption{Snapshots of $2d$ configurations of the particle model at
  different times. In all panels, parameters are $N=10^4$ and
  $D=10^{-4}$. Details of the particle simulations are in
  \cite{suppl}. (left column) The two species
  have the same diffusivity. (right column) The red species
  has diffusivity $D+\delta D$ with $\delta D/D=0.05$. To help the
  viewer, configurations has been downsampled (one every four
  particles, chosen at random, is shown). \label{fig_ill}}
\end{center}
\end{figure}

We consider a model in which particles belonging to two
different species $A$ and $B$ diffuse in space, reproduce according to
the reactions $A\rightarrow 2A$ and $B\rightarrow 2B$, and die in a
density-dependent fashion ($A+A/B\rightarrow A$ and $B+A/B\rightarrow
A$) as result of competition. For simplicity, we assume all reactions
to occur at the same rate $\mu=1$. The system is an hypercube of size
$L^d$ in $d$ dimensions with periodic boundary conditions (see
\cite{longpaper,suppl} for details on the implementation). We call
$D+\delta D$ and $D$ the diffusion constants of species $A$ and $B$
respectively. When $\delta D=0$, the dynamics of the model is
characterized by a coarsening process, as shown in the two-dimensional
example of Fig. (\ref{fig_ill}), left column. We will later argue that
this coarsening belongs to the universality class of the VM. Instead,
Fig. (\ref{fig_ill}), right column shows that a small difference in
the diffusivity of the two species, $\delta D/D= 5\%$ in this case,
imposes a non-negligible bias on the coarsening dynamics and, in
particular, confers to an advantage to the species having a larger
diffusivity. A similar behavior can be observed also in 1d
simulations.

We derived macroscopic equations for the concentrations of the two species  $c_A(\mathbf{x},t)$ and
$c_B(\mathbf{x},t)$ 
\cite{gardiner,prl_pigolotti,longpaper,suppl}. The result is
\begin{eqnarray}\label{model}
\partial_t c_A &=& (D+\delta D)\nabla^2 c_A+\mu c_A(1-c_A-c_B)+
\sigma_A\xi_A 
\nonumber\\
\partial_t c_B  &=& D\nabla^2 c_B+\mu c_B(1-c_A-c_B)+
\sigma_B\xi_B
\end{eqnarray}
where $\sigma_A=\sqrt{\mu c_A(1+c_A+c_B)/N}$ and similarly for
$\sigma_B$. The parameter $\mu/N$ is the genetic drift and $N$ can be
interpreted as the density of particles corresponding to a macroscopic
concentration $c=1$.  $\xi_A(\mathbf{x},t)$ and $\xi_B(\mathbf{x},t)$
are independent delta-correlated (in space and time) noise sources,
$\langle
\xi_i(\mathbf{x},t)\xi_j(\mathbf{x'},t')\rangle=\delta_{ij}\delta(\mathbf{x}-\mathbf{x'})\delta(t-t')$.
An equation for the relative concentration of one species $f \equiv
c_A/(c_A+c_B)$ can be obtained from Eqs. (\ref{model}) by means of
Ito's formula \cite{prl_pigolotti,longpaper}. Performing the
calculation and neglecting fluctuations of the total particle density
by imposing $c_A+c_B=1$ at the end of the procedure yields
\begin{equation}
\label{simple}
\partial_t f (\mathbf{x},t)= D \nabla^2 f + \delta D (1-f)\nabla^2 f
+ \sigma \xi (\mathbf{x},t)
\end{equation}
where $\sigma= \sqrt{ 2\mu f(1-f)/N}$. Eq. (\ref{simple}) constitutes
the starting point of our analysis.  We remark that, while we
neglected fluctuations of the total density $c_A+c_B$, the fact that
the total density is not strictly conserved is crucial to derive
Eq. (\ref{simple}). This correspond to the fact that it is impossible
to have two species with different diffusion constant in a lattice
model: if each site is strictly occupied by one spin, then the
effective diffusivity of the two species is equal by constraint.
Setting $\delta D=0$ in (\ref{simple}) one retrieves Eq. (\ref{VM})
describing the VM universality class \cite{longpaper,revmodph,hammal}.
Although (\ref{simple}) has been derived thinking of the continuum
limit of a biological model, we argue that its validity is more
general, as the term proportional to $\delta D$ is the simplest,
non-trivial way to account for a difference in diffusivity between the
two species. In the following, we will study how this term affects
the dynamics by breaking the the VM universality class.

\begin{figure}[htb]
\begin{center}
\includegraphics[width=8.4cm]{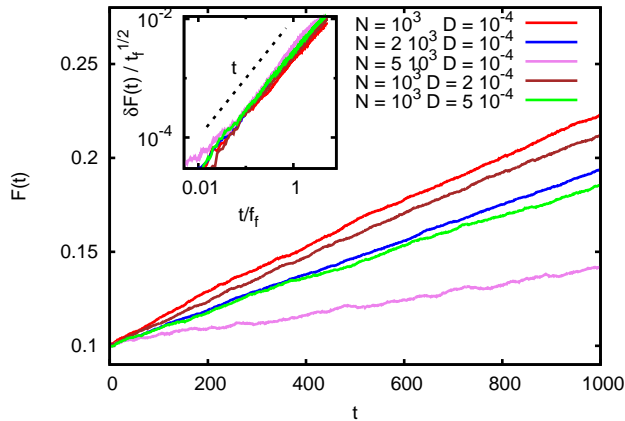}
\caption{Behavior of the average concentration  of the fast species $F(t)$ ar varying
  the number density $N$ and the diffusion constant of the slowest
  species $D$ in one dimension. In all simulations, the relative difference of diffusion
  constants is $\delta D/D=0.1$, the initial fraction of the fast
  species is $F(0)=0.1$ and the system size is $L=10$. Curves
  are average over $10^3$ realizations. The inset shows a
  data collapse according to Eq. (\ref{universal}). Linear scaling
  (black dashed line) is  shown for comparison. 
  \label{fig2}}
\end{center}
\end{figure}

We start by
focusing on the $1d$ case and study the time evolution of the integrated
mean concentration $F(t) = \langle f \rangle $, where $\langle
.. \rangle$ denotes an average over space and noise.  From
Eq. (\ref{simple}) we obtain
\begin{equation}
\frac{dF}{dt} = \delta D\ \langle (\nabla f)^2 \rangle > 0
\end{equation}
which is Eq. (\ref{naive}). The above equation already shows that
$F(t)$ is a growing function of time for any $\delta D >0$. The
behavior of $F(t)$ is presented in Fig. (\ref{fig2}) in $1d$
simulations of the particle model, starting with uniformly distributed
populations but a more abundant slow species, so that $F(0)=0.1$.
Notice how $F(t)$ decreases at increasing $N$ and increasing $D$ at
constant $\delta D/D$. A straightforward calculation \cite{suppl}
shows that
\begin{equation}\label{hethet}
  \frac{dF(t)}{dt} = \frac{\delta D}{2} \nabla^2 H (x,t) |_{x=0} 
\end{equation}
where we introduced the two point connected correlation function ({\em
  heterozygosity} in biological language) $H(x = x_1-x_2,t) = \langle
f(x_1)[1-f(x_2)]+f(x_2)[1-f(x_1)] \rangle$, which is function of
$x_1-x_2$ only due to translational invariance. For $\delta D=0$ the
function $H(x,t)$ is explicitly known \cite{revmodph}. For $\delta
D/D\ll1$, we can use this result to evaluate the right hand side of
Eq. (\ref{hethet}) perturbatively , i.e. by replacing the average
$\langle \dots \rangle$ with the average $\langle \dots \rangle_0$
over the solvable case of $\delta D=0$ \cite{suppl}. The result is
\begin{equation}
\frac{dF(t)}{ dt} =  \frac{ \delta D}{4D\sqrt{\pi \epsilon t_f} } H(0)  {G(t/t_f)} 
\label{integral3}
\end{equation}
where $G(x)=\exp(x)\mathrm{erfc}(\sqrt{x})$, $t_f=2DN^2$ and the
parameter $\epsilon$ is an ultraviolet cutoff that can be assumed to
be of order 1.
Using expression (\ref{integral3}), we can cast the growth of
$\delta F(t)=F(t)-F(0)$ into the scaling form
\begin{equation}
\delta F(t)=H(0)\sqrt{t_f}\frac{\delta D}{D}\phi(t/t_f) 
\label{universal}
\end{equation}
where the scaling function $\phi$ does not depend on parameters and is
$ \phi(x) \sim x$ for small $x$. A collapse of curves for
different values of $N$ and $D$ according to (\ref{universal}),
presented in the inset of Fig. (\ref{fig2}), supports our theory
within statistical fluctuations. Notice that, at this order in
perturbation theory, the presence of absorbing states is not predicted
by Eq. (\ref{integral3}). This means that the perturbative approach is
expected to describe properly the dynamics only on times shorter than
the global fixation time, i..e. $L^2/D$.

It is of interest to compare a difference in diffusivity to a
selective advantage caused by a difference $s$ in reproduction rates, i.e. in the case of
equation (\ref{FKPP}).
Assuming $s\ll 1$ and averaging directly such
term, one obtains that $F(t)$ evolves in this case according to $dF/dt
=sH(0)G(t/t_f)/2$. Comparing the latter expression with
Eq. (\ref{integral3}), it is natural to define an {\it
  effective} advantage given by
\begin{equation}
s_{eff}   =  \frac{\delta D}{ 2 D \sqrt{ \epsilon \pi t_f }}.
\label{vantaggio}
\end{equation}

To tackle the problem in higher dimensions, let us start from the
general evolution equation for the two point connected correlation function  as obtained
from Eq. (\ref{simple}) for $\delta D=0$:
\begin{equation}
\partial_t H(x,t) = 2D \nabla^2 H - \frac{2\mu}{N} H(0,t) \delta(x).
\label{Hxt}
\end{equation}
Due to the spatial regularization \cite{doering}, the delta
  function resulting from Ito calculus must be
  interpreted as $\delta(x) \sim 1/a^d$, where
$a \sim \sqrt{2D \epsilon}$ is the lattice spacing of the discrete
stepping stone model. In an adiabatic approximation of
Eq. (\ref{Hxt}), $\nabla^2 H|_{x=0}$ can be estimated as
\begin{equation}
  \nabla^2 H|_{x=0}  \sim  \frac{\mu H(0,t)}{DN  (D\epsilon)^{d/2}} 
\label{deltaH}
\end{equation}
which is consistent with the scaling of Eq. (\ref{universal}) for
$d=1$.  Evaluating Eq. (\ref{deltaH}) in $d=2$ yields $\nabla^2
H|_{x=0} \sim \mu H(0,t) / (ND^2 \epsilon) $, i.e. the effective
advantage becomes larger by a factor $1/\sqrt{D\epsilon}$ with respect
to the one dimensional case.

\begin{figure}[htb]
\begin{center}
\includegraphics[width=8.7cm]{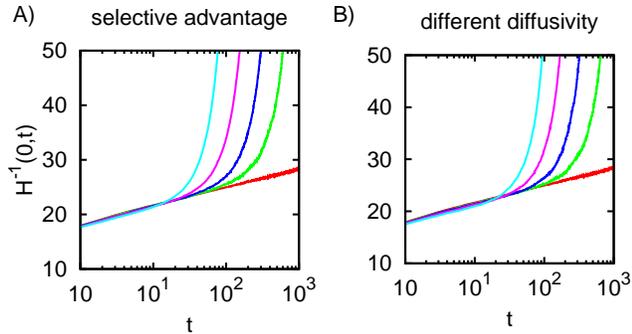}
\caption{Comparison of the coarsening dynamics in $2D$ in the presence
  of A) selective advantage (i.e. faster reproduction rate of species
  A, and B) difference in diffusivity. Parameters
  are $D= 2\cdot 10^{-5}$ and $N=4\cdot 10^4$; the system is a square
  of side $L=1$. In A), the selective advantage $s=0$ (red curve),
  $s=0.05, 0.01, 0.02, 0.04$ (curves from right to left). In B) we
  vary $\delta D/D =0$ (red curve), $\delta D/D=0.04, 0.08, 0.16,
  0.32$ (curves from right to left). In both panels, we plot the
  inverse density of interfaces $H^{-1}(x=0,t)$ as a function of
  time. Notice how the logarithmic behaviour, characteristic of the VM
  is cut off on a time set by either the selective advantage or the
  difference in diffusivity.
  \label{fig4}}
\end{center}
\end{figure}

The interpretation of Eqs. (\ref{vantaggio}) and (\ref{deltaH}) is
that, after averaging over noise and space, the effect of a different
diffusivity is analogous to that of a selective advantage. We now
discuss the consequences for the peculiar coarsening properties of the
VM. In $2D$, the dynamics of the VM is characterized by a slow
coarsening process where the density of interface decays as
$\log^{-1}(t)$ (see e.g. \cite{dornic}). In the continuum off-lattice
case, the analogous of the density of interface is the local
heterozigosity $H(x=0,t)$.  Fig. (\ref{fig4}) shows how $H(x=0,t)$
displays the expected logarithmic decay in our particle model. When either a
selective advantage or a diffusivity difference is present, this
behavior is observed up to a time $\bar{t}$ (either proportional to  the
selective advantage $s$ or the diffusivity difference $\delta D$)
after which $H(0,t)$ decays exponentially. This shows how both
terms have a similar effect in driving the dynamics away from the VM
critical point.

\begin{figure}[htb]
\begin{center}
\includegraphics[width=9cm]{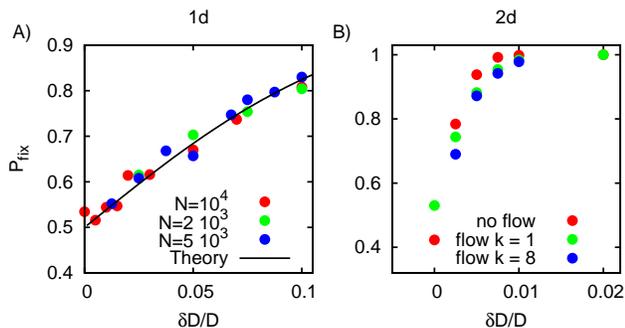}
\caption{(A) Fixation
  probabilities in 1d. The
  black line is the theoretical prediction of Eq. (\ref{kimdoer}). (B)
 Fixation probabilities in 2d, with and without fluid
 flow. In both panels  we set  $D=10^{-4}$. \label{fig3}}
\end{center}
\end{figure}

The effective selective advantage introduced in Eq. (\ref{vantaggio})
can be used to study the probability of fixation $P_{fix}$, defined as
the probability of reaching the absorbing state $f=1$ of Eq. (\ref{simple}).
The fixation probability in terms of a selective
advantage is given by the formula \cite{doering,longpaper}
\begin{equation}\label{kimdoer}
 P_{fix}(s) = \frac{1-\exp(-2sNF(0))}{1-\exp(-2sN)}.
\end{equation}
Assuming the same formula to hold in the case of different
diffusivities with $s_{eff}$ replacing $s$ leads to an interesting
prediction: $P_{fix}$ should not depend on $N$ as $s_{eff}\propto
N^{-1}$.  In Fig. (\ref{fig3}A) we show $P_{fix}$ as a function
of $\delta D /D$ for different values of $N$, confirming this
prediction. The black line is the expression for $P_{fix}$, namely:
obtained from $\epsilon \approx 0.5\mu^{-1}$.  This also confirms our
initial assumption of $\epsilon\mu \sim 1$. Notice how the bias in the fixation
probabilities shown in Fig. (\ref{fig3}) is much stronger in $2d$
than in $1d$ at equal parameter values, as predicted by Eq. (\ref{deltaH})

We now briefly discuss the same problem in the presence of advection.
Simulations in 2D shown in Fig. (\ref{fig3}B) show similar fixation
probabilities with and without advection by an incompressible flow
(details in \cite{suppl}).  In \cite{suppl} we argue that in an
incompressible flow Eq. (\ref{naive}) formally holds, leading to the
same effective advantage for the species with larger diffusivity, as
far as the typical scale of turbulent scale is not too small.

To conclude, we have shown that a small difference in diffusivity can
induce a breaking of the VM universality class with the critical
parameter proportional to the noise amplitude. In the framework of
population dynamics, this means that a difference in diffusivity
between two species can bias the outcome of competition towards the
more agile one. Notice that, while in the presence of a range
expansion there exist an advantage for the fastest species which can
be estimated by looking at the difference between the deterministic
Fisher velocities \cite{korolev_sweeps,benichou}, the effect presented
here is genuinely stochastic and constitutes a new example of a
noise-induced advantage in population genetics
\cite{hall_noise,heinsalu}.

When considering a realistic biological scenario, the effective
advantage given by a higher motility must be compared with its
involved metabolic cost. In this respect, our result is reminiscent of
a classic analysis of seed dispersal by Hamilton and May
\cite{hamilton1,hamilton2}, which predicts an equilibrium, optimal
level of dispersal even in an homogeneous environment.  Further, our
preliminary results in the presence of fluid flows suggest that the
same effect can be crucial also in turbulent marine environments.  We
expect our result to be also relevant in other fields where diffusion
is known to affect crucially the dynamics, such as chemical kinetics
\cite{krap}, game theory \cite{frey}, and synchronization
\cite{diaz}. Indeed, a term proportional to $(1-f)\nabla^2 f$ would
generally appear in the continuous description of systems
characterized by two spatial concentrations diffusing at different
speed, the proper dynamical evolution being described by
Eq. (\ref{simple}).
  
\section{Supplementary Material}

\section{Particle model and macroscopic equations}\label{sec1}

\subsection{Particle model}

The particle model is implemented as follow. We consider
particle belonging to two different species, $A$ and $B$. All
particles diffuse in space, with diffusion constants equal to $D+\delta
D$ for species $A$ and $D$ for species $B$. In the final part of the
Letter, we consider also advection by a velocity field, which simply
transports particles as Lagrangian passive scalars.  Integration of
the particle motion is performed via a second-order Adams-Bashforth
scheme.

Superimposed to particle movement, we implemented the stochastic
reactions corresponding to reproduction and death of individuals. All
possible reactions are listed in the following table

\begin{table}[htb]
\begin{tabular}{|p{5cm}|*{2}{c|}|}
     \hline
birth & death \\
\hline
$A \rightarrow 2A$ $(\mu_A)$& $A+A\rightarrow A$ $(\tilde{\lambda}_{AA})$ \\
                                & $A+B\rightarrow A$ $(\tilde{\lambda}_{AB})$\\
\hline
$B \rightarrow 2B$ $(\mu_B$) & $B+A\rightarrow B$ $(\tilde{\lambda}_{BA})$\\
                                & $B+B\rightarrow B$ $(\tilde{\lambda}_{BB})$\\
\hline
\end{tabular}
\end{table}

where in parenthesis we denote the rate at which each reaction takes
place, which is a free parameter.  Binary reactions corresponding to
death events are implemented locally: the rate is proportional to the
number of individual of the corresponding type in a hypercube of
volume $\delta$ centered on the considered individual ($\delta$ has
dimensions of a length to the power $d$ in $d$ dimensions).  Reactions
are numerically implemented with a simple stochastic first-order
scheme.  At each time step $dt$, with probability $\mu_A
dt$ an individual of species $A$ reproduces, while with probability
$(\lambda_{AA}n_A+\lambda_{AB}n_B)dt$ it dies, being $n_A$ and $n_B$ the
number of neighbors of type $A$ and $B$ defined via the hypercube
described above. Finally, newborn individuals are initially placed at the
same coordinate of their mother.

\subsection{Macroscopic equations}

There exist several methods to derive macroscopic equations for
particle models \cite{gardiner,risken,vankampen}. We adopted the
procedure of the ``chemical master equation'', as detailed in
\cite{gardiner}, chapter 8 and \cite{longpaper}.

Let us divide the systems in cells of size $a$ (where $a$ is a length
to the power $d$ in $d$ dimensions). The parameter $a$ should be
sufficiently large so that the average number of particles inside each
cell is sufficiently large, yet small enough so that the empirical
density of particle does not change appreciably inside the
cell.  Under these assumptions, we now compute the rates
$W_{A,B}(\pm,n^A_j,n^B_j)$ at which the population of species $A$($B$)
in a given cell labeled by index $j$ increases/decreases of one unit
given the current values of the populations. Such rates read

\begin{eqnarray}\label{mrates}
  W_A(+1,n_j^A,n_j^B)&=&\mu_A n_j^A +\nonumber\\
  &+&\mathrm{D/A}\nonumber\\
  W_A(-1,n_j^A,n_j^B)&=&\frac{\delta}{a}(\tilde{\lambda}_{AA}n_j^A(n_j^A-1)+\tilde{\lambda}_{AB}n_j^An_j^B)+\nonumber\\
  &+&\mathrm{D/A}\nonumber\\
  W_B(+1,n_j^A,n_j^B)&=&\mu_A n_j^B +\nonumber\\&+&\mathrm{D/A}\nonumber\\
  W_B(-1,n_j^A,n_j^B)&=&\frac{\delta}{a}(\tilde{\lambda}_{BA}n_j^An_j^B+\tilde{\lambda}_{BB}n_j^B(n_j^B-1))+
\nonumber\\
  &+&\mathrm{D/A}.
\end{eqnarray}

where by ``D/A we mean the contributions due to diffusion and
advection of particles among neighboring cells that we did not write explicitly.
Such equations allow for writing the master equations governing the
probability of having $n_j^A,n_j^B$ particles of species $A$ and $B$
in each cell $j$. The chemical master equation method \cite{gardiner}
consists in performing a Kramers-Moyal expansion of such master
equation in each of these cells. We recall that the Kramers-Moyal
method works as follows. Let us call $W(\Delta n, n)$ the transition
rate from $n$ particles to $n+\Delta n$ particles in a master
equation. Performing a Taylor series expansion of the master equation
in $\Delta n$ yields
\begin{equation}\label{KM1}
  \partial_t P(n,t)=\sum\limits_{j=1}^{\infty}
\frac{(-1)^j}{j!}\partial^j_n[\alpha_j(n)P(n,t)]
\end{equation}
with
\begin{equation}\label{KM2}
\alpha_j(n)=\int d\Delta n \ (\Delta n)^j W(\Delta n,n).
\end{equation}
Finally, truncation of Eq. (\ref{KM1}) at the second order in $j$
leads to a Fokker-Planck equation.  We now apply this procedure to
each cell in our case. It is convenient to introduce the macroscopic
binary rates $\lambda_{lk}=N\delta\tilde{\lambda_{lk}}$, where the
constant parameter $N$ has the dimension of a density and represents
the typical number of particles per unit volume. For convenience, we
write the equivalent Langevin equation instead of the Fokker-Planck
\begin{eqnarray}
  \frac{dn^A_j}{dt}&=&n^A_j(\mu-\lambda_{AA}n^A_j/Na-\lambda_{AB}n^B_j/Na)
+\mathrm{D/A}+\nonumber\\
&+&\sqrt{n^A_j(\mu+\lambda_{AA}n^A_j/Na+\lambda_{AB}n^B_j/Na)}\xi^i_A
  \nonumber\\
  \frac{dn_B^i}{dt}&=&n_B(\mu-\lambda_{BA}n^A_j/Na-\lambda_{BB}n^B_j/Na)
+\mathrm{D/A}+\nonumber\\
  &+&\sqrt{n^B_j(\mu+\lambda_{BA}n^A_j/Na+\lambda_{BB}n^B_j/Na)}\xi^i_B.
\end{eqnarray}
The noise terms obey
$\langle \xi^i_{k}(t)\xi^j_{k'}(t')\rangle
=\delta(t-t')\delta_{j,j'}\delta_{k,k'}$ and must be interpreted
according to the Ito prescription. In principle, diffusion and
advection terms affect also the noise amplitude. However, one can show
that their effect can be neglected if the cell size $a$ is chosen
sufficiently large \cite{gardiner}.

We now  introduce macroscopic concentrations $c^{A,B}_j=n^{A,B}_j/(Na)$. The Langevin equations for the
concentrations read
\begin{eqnarray}
  \frac{dc^A_j}{dt}&=&c^A_j(\mu-\lambda_{AA}c^A_j-\lambda_{AB}c^B_j)
+\mathrm{D/A}+\nonumber\\
&+&\sqrt{c^A_j(\mu+\lambda_{AA}c^A_j+\lambda_{AB}c^B_j)}\frac{\xi^i_A}{\sqrt{aN}}
  \nonumber\\
  \frac{dc^B_j}{dt}&=&c^B_j(\mu-\lambda_{BA}c^A_j-\lambda_{BB}c^B_j)
+\mathrm{D/A}+\nonumber\\
  &+&\sqrt{c^B_j(\mu+\lambda_{BA}c^A_j+\lambda_{BB}c^B_j)}\frac{\xi^i_B}{\sqrt{aN}}
\end{eqnarray}

It is now possible to write the continuum equation by formally taking
the limit $a\rightarrow 0$
\begin{eqnarray}\label{cont_macr}
  \frac{dc^A}{dt}&=&c^A_j(\mu_A-\lambda_{AA}c^A-\lambda_{AB}c^B)
+(D+\delta D)\nabla^2 c_A \nonumber\\
&-&\nabla (vc_A)
+\sigma_A \xi(x,t)
  \nonumber\\
  \frac{dc^B}{dt}&=&c^B_j(\mu_A-\lambda_{BA}c^A-\lambda_{BB}c^B)
+D\nabla^2 c_B \nonumber\\
&-&\nabla (vc_B)
  +\sigma_B\xi'(x,t)
\end{eqnarray}
where we wrote explicitly the macroscopic deterministic expression fo
the advection and diffusion terms ($v$ is the advecting velocity). The noise terms now obey
$<\xi(x,t)\xi'(x',t')>=0$,
$<\xi(x,t)\xi(x',t')>=\delta(t-t')\delta(x-x')$ and
$<\xi'(x,t)\xi'(x',t')>=\delta(t-t')\delta(x-x')$ . We also defined
$\sigma_A=\sqrt{\mu c_A(1+c_A+c_B)/N}$ and $\sigma_B=\sqrt{\mu
  c_B(1+c_A+c_B)/N}$. Finally, Eq. (3) is obtained
from Eq. (\ref{cont_macr}) by setting all rates equal,
$\mu_i=\lambda_{ij}=\mu$ $\forall i,j$. This choice corresponds to the
biological case in which the two species are neutral variants of each
other and the carrying capacity for the concentrations is set to $1$
(see \cite{longpaper} for the general case in which these rates are allowed
for being different).

We remind that the continuous limit of such
reaction-diffusion-advection processes is only a concise description of
the stochastic dynamics, which is mathematically well-defined only in
the space-discretized version (see Ref. \cite{gardiner}, page 314 and
the discussion in \cite{longpaper}). Indeed, we shall see in Section 3
of this supplements an example of a
property of the macroscopic equation which depends
explicitly on the choice of the lattice spacing $a$.

\section{Perturbative expansion}

In this section, we provide details of the derivation leading to
Eq. (7) of the Letter.  We start from (Eq. \ref{simple}).
As argued, the
mean concentration $F(t) = \langle f \rangle $, where $\langle
.. \rangle$ denotes an average over space and noise, evolves according
to
\begin{equation}
\label{naive}
\frac{dF}{dt} = \delta D\ \langle (\nabla f)^2 \rangle > 0 .
\end{equation}
Let us now define the quantity $h(x_1,x_2,t) \equiv
f(x_1)[1-f(x_2)]+ f(x_2)[1-f(x_1)]$.  A simple computation shows that
\begin{equation}
\label{identity}
\nabla^2_{x_1} f(x_1)+
\nabla^2_{x_1} h (x_1,x_2,t) =
2 \nabla^2_{x_1} f(x_1) [1-f(x_2)]
\end{equation}
We now introduce the {\em heterozygosity} $H(x = x_1-x_2,t) = \langle
h(x_1,x_2,t) \rangle$, which is function of $x=x_1-x_2$ as argued in
the main text. Upon using (\ref{identity}) and reminding
that $\langle \nabla^2  f(x) \rangle = 0$ we obtain from
Eq. (\ref{naive})
\begin{equation}
  \frac{dF(t)}{dt} = \frac{\delta D}{2} \nabla^2 H (x,t) |_{x=0}
\end{equation}
For $\delta D=0$, the function $H(x,t)$ can be explicitly computed:
\cite{revmodph}:
\begin{equation}
\label{hetero}
\frac{H(x,t)}{H(0)}  = 1 -   \frac{2}{N} \int_0^t ds  \frac{e ^{- {x^2 \over 8D(t-s)}}}{\sqrt{8\pi D(t-s)}} G(s/t_f)
\end{equation}
where $t_f \equiv 2DN^2$ and $G(x) = \exp(x)\
\mathrm{erfc}(\sqrt{x})$. Our aim is now to evaluate the growth of $F(t)$ in a perturbative way,
i.e. by approximating the average above with the average
$\langle\dots\rangle_0$ on the solvable version of Eq. (\ref{simple})
with $\delta D=0$. Assuming $\delta D/D \ll 1$ we can use
$H(0,x,t)$ given in (\ref{hetero}) and obtain:
\begin{equation}
\frac{dF(t)}{ dt} = \frac{\delta D}{8 D\sqrt{\pi}t_f} \int_0^{\tau}
\frac{H(0) G(s)\ ds}{(\tau-s)^{3/2}}
\label{result}
\end{equation}
where $\tau=t/t_f$.  Within the approximation mentioned above
(i.e. $\delta D/D \ll 1$, $C_T \approx 1$), equation (\ref{result}) is
the solution of our problem in $1d$. However, the integral in
Eq. (\ref{result}) diverges for $s\rightarrow \tau$. To avoid this
singularity, one must introduce a cutoff time $\epsilon$ and rewrite
Eq. (\ref{result}) in the form
\begin{eqnarray}
\frac{dF(t)}{ dt} = \frac{\delta D}{8 D\sqrt{\pi}t_f}
\int_0^{\tau-\epsilon_f} \frac{ H(0) G(s)\ ds}{ [\tau-s]^{3/2} }
\label{cutoff}
\end{eqnarray}
where $\epsilon_f \equiv \epsilon/t_f$. An explicit computation of the
integral in (\ref{cutoff}) gives:
\begin{equation}
\frac{dF(t)}{ dt} =  \frac{\delta D}{8 D\sqrt{\pi}t_f}  \frac{H(0)g(t/t_f)}{\sqrt{\epsilon_f}}
\label{integral}
\end{equation}
where
\begin{equation}
g(\tau) = 2 G(\tau) + O(\epsilon_f^{1/2}).
\label{integral2}
\end{equation}
One can conjecture that the cutoff $\epsilon$ should be on the order
of the shortest timescale in the system, i.e. by the reaction
rates. Under this timescale, we should expect a breakdown of the
macroscopic description due to the discrete nature of birth/death
events.  This argument suggests that $\epsilon\mu \approx 1$, implying
$\epsilon_f\ll1$.  Substituting (\ref{integral2}) into
(\ref{integral}) yields, to the leading order, to Eq. (7) of the Letter.

\section{Cole-Hopf approach}

In this section, we briefly discuss an alternative approach to the
perturbative expansion based on a Cole-Hopf transformation, in analogy
with \cite{hall_noise}. To shorten the notation, let us define the small
parameter $\eta=\delta D/D$. Let us start from Eq. (\ref{simple}) and
make the Cole-Hopf change of variable
\begin{equation}
f(x,t) \ \rightarrow g(x,t)=D\frac{e^{2\delta D f(x,t)/D}-1}{2\delta D}.
\end{equation}
The equation for $g(x,t)$ can be simply derived by means of Ito's
formula. At the first order in $\delta D$ and after averaging over the
noise, it reads
\begin{equation}\label{colehopf_res}
\partial_t g + \delta D \partial_x (g \partial_x g)= (D+\delta D)
\Delta g + \frac{\delta D}{2D N a} g(1-g)
\end{equation}
where the last term is a Ito term. The parameter $a$ appearing in the
denominator of such term is the spatial mesh introduced in Section
1. As usually happen in spatially-extended Langevin equation, the
resolution of the spatial mesh appears explicitly when performing a
nonlinear change of variable (see also \cite{doering}). This is
related to the ill-defined properties of the continuum limit also
discussed in Section 1. Finally, note that the necessity of specifying
the size of the mesh in this approach corresponds exactly to the
necessity of introducing an ultraviolet cutoff in the approach
followed in Section 2.

Upon averaging Eq. (\ref{colehopf_res}) in space, we obtain:
\begin{equation}
\label{itof}
\frac{d}{dt} \langle g  \rangle \sim  \frac{\delta D}{ 2D N a} \langle g(1-g) \rangle.
\end{equation}
Let us notice that $f=g$ up to order $\delta D/D$.

As briefly discussed in the manuscript after Eq. (10), we can relate the {\em
  spatial} cutoff $a$ and the {\em time} cutoff $\epsilon$ via the relation
$a \sim (D \epsilon)^{1/2}$. Notice that this is also consistent with the
scaling assumed in the continuum limit of the Stepping Stone model,
see e.g. \cite{doering,revmodph}.  Finally, we argue that $\epsilon$ is a
fraction of the time scale $1/\mu$ since it is the shortest time scale
in the system assuming $1/\mu \ll L^2/D$.  Putting everything
together, we obtain:
\begin{equation}
\label{finale}
\frac{d}{dt} \langle f \rangle \sim  \frac{\delta D}{ 2DN \sqrt { D\epsilon} } \langle f(1-f) \rangle
\end{equation}
which is consistent with Eq.  (7) in the manuscript.

\section{Effect of fluid advection}

\begin{figure}[htb]
\begin{center}
\includegraphics[width=10cm]{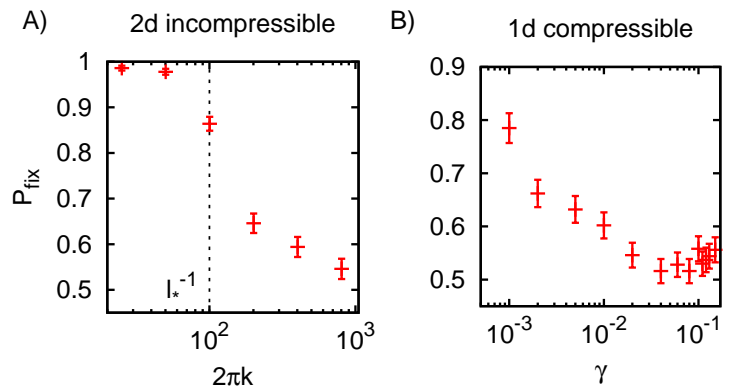}
\caption{(A) Fixation probability in the two-dimensional
  incompressible flow of Eq. (\ref{eqflow}) as a function of the
  inverse turbulent scale $k$. Parameters are: $\gamma=1$, $\delta
  D/D=10^{-2}$, $N=10^4$, $D=10^{-4}$ and $L=1$. For these parameters,
  one has $l_*\approx 0.01$. (B) Fixation probability in a
  one-dimensional compressible flow $v=\gamma \sin(2\pi x)$, as a
  function of the forcing intensity $\gamma$. Parameters are $D=2\cdot
  10^{-5}$, $N=10^3$, $\delta D/D=0.2$, $f_0=1/2$, $L=1$.\label{fig5}}
\end{center}
\end{figure}

In this section, we briefly discuss the effect of different
diffusivities in the presence of an advecting
velocity field ${\bf v}$.  In this case, the macroscopic description
of Eqs. (3) in the main text is still valid upon replacing $\partial_t$ with
a material derivate, i.e.  $\partial_t \rightarrow \partial_t + div(
{\bf v} f)$.  We first consider a $2d$ time-dependent incompressible
flow leading to Lagrangian chaos
\begin{eqnarray}
v_x (x, y, t)&=&\gamma[\cos(2k\pi y)+\sin(2k\pi y)\cos(t)]  \nonumber\\
v_y (x, y, t)&=& \gamma[\cos(2k\pi x)+\sin(2k\pi x)\cos(t)]. \label{eqflow}
\end{eqnarray}
In the presence of advection, one may be tempted to formulate the
problem by replacing the diffusion constant $D$ with the so called
"eddy diffusivity" $D_{turb} \sim u_* l_{turb}$ where $u_*$ is r.m.s
velocity and $l_{turb}$ is the turbulent mixing length of the flow.
For the flow of Eq. {\ref{eqflow}}, it has been shown that in a wide
range of $D$, $D_{turb}$ is practically independent of $D$
\cite{biferale}. Therefore, one may reach the conclusion that a small
difference in the bare diffusivity should hardly have any effect.
However, simulations in Fig. (3B) in the main text show similar fixation
probabilities with and without advection.  In fact, in an
incompressible flow, Eq. (4) in the main text still holds, leading to the
effective advantage for the species with larger diffusivity. Notice
however that Eq. (9) in the main text naturally identifies a spatial
scale $l_*^2 = DN(D\epsilon)^{d/2}/\mu$, independent of $\delta D$.
For the flow of Eq. (\ref{eqflow}) one can assume $l_{turb} \approx
(2\pi k)^{-1}$.  When $l_{turb}$ is much larger than $l_*$, turbulence
should not alter the effective advantage of the allele with larger
diffusivity.  In the opposite limit of $l_{turb}\ll l_*$, we expect
that turbulence should disrupt the effect.  In Fig. (\ref{fig5}A), we
test this idea by studying how the fixation probability at $\delta
D/D=0.1$ changes at varying $k$.  Upon decreasing the forcing scale
$k^{-1}$ much below the scale $l_*$, the bias in the fixation
probabilities vanishes as expected.

To assess the relevance of this result for competition in the ocean,
we consider phytoplankton dynamics in the oceanic upper layer.
Typical maximum phytoplankton densities set by nutrient concentrations
in the ocean are $N\sim 10^8 cells/m^3$, \cite{carrcap}, while
average diffusivities are estimated to be $D\sim 10^{-9} m^2/s$
(\cite{dataplankton}). In $3d$, this leads to an estimate of $l_* \sim
0.1 m$. In the upper oceanic layer, a suitable measure of $l_{turb}$
comes from measured values of the turbulent mixing
length used to parameterize momentum and heat transport. According to
\cite{turboocean}, the estimate of $l_{turb}$ is in the range $1-10$
meters, i.e. much larger than $l_*$. Although these numbers may of
course vary considerably depending on the conditions, we argue that
the condition $l_* < l_{turb}$ is likely to be satisfied in most
marine environments.

Finally, we briefly discuss what happens in the presence of a
compressible flow, where species localize near sinks in the velocity
field \cite{prasad} \cite{prl_pigolotti}. The characteristic
localization scale $l_c$ near a sink is of order $\sqrt{D/\Gamma }$
where $\Gamma$ is the typical velocity gradient \cite{benzi}. When
$l_c$ is smaller than $l_*$, the effect due to localization should
greatly reduce the effect of having a larger diffusivity.
Consequently, we expect a significant decrease in $P_{fix}$ for
$\Gamma \sim \sqrt{\mu/t_f}$. This scenario is supported by
simulations shown in Fig. (\ref{fig5}B), where we plot $P_{fix}$
obtained in a $1d$ system subject to a simple compressible flow,
$v(x)=\gamma \sin(2\pi x)$, see \cite{prl_pigolotti} for a comparison
with $\delta D=0$. As predicted, we observe a strong decrease of
$P_{fix}$ at increasing $\gamma$.

  \begin{acknowledgments}
  
  We wish to thank M. Cencini, D.R. Nelson and M.A. Mu\~{n}oz for
  comments on the manuscript.  SP acknowledges partial support from
  Spanish research ministry through grant FIS2012-37655-C02-01.
  
\end{acknowledgments}

\end{document}